\documentclass[prb,preprintnumbers,superscriptaddress,amsmath,amssymb,floatfix,aps,twocolumn]{revtex4}
\usepackage{graphicx}
\usepackage{xcolor}
\usepackage{lipsum} 
\usepackage{xcolor}
\usepackage{afterpage}

\usepackage[version=4]{mhchem}
\usepackage{amsmath}
\usepackage{amsbsy}
\usepackage{float}
\usepackage[utf8]{inputenc}
\usepackage{tabularx}
\usepackage{booktabs}
\usepackage{array}
\usepackage{multirow}
\usepackage{adjustbox} 
\usepackage{float}
\usepackage{tikz}
\usepackage{tikz-3dplot}
\usepackage{hyperref}
\usepackage{cleveref}

\begin{document}

\title{High-temperature superconducting Majorana fermions platforms in the layered Kitaev Materials: Case study of $Li_2IrO_3$}

\date{\today} 

\author{Elnaz Rostampour}
\author{Badie Ghavami}
\email{ghavamiba@gmail.com}
\affiliation{Department of Quantum Materials,  Qlogy Lab Inc.}

\begin{abstract}
Recent advances in Kitaev materials have highlighted their potential to host Majorana fermions without or high-temperature of superconductivity.  In this research, we propose $Li_2IrO_3$ as a promising High-temperature superconducting platform supporting Majorana edge modes due to its strong spin-orbit coupling, honeycomb  lattice structure, and proximity to a quantum spin liquid (QSL) phase. A theoretical and numerical framework based on the Kitaev-Heisenberg Hamiltonian is developed to model spin interactions in $Li_2IrO_3$.
Here, the existence of topological zero-energy states is demonstrated, and their signatures in the edge-localized spectral weight are identified.
A device concept based on this material is also proposed with potential industrial applications in spintronics, magnetic field sensing, and topological quantum memory.

\end{abstract}

\maketitle
 
\section{\label{sec:intro} Introduction}
Majorana fermions have become a focal point in condensed matter physics and quantum materials due to their non-abelian exchange statistics and potential role in fault-tolerant quantum computation \cite{beenakker2013search,leijnse2012introduction,stanescu2013majorana,kozii2016three}. While most proposals for their realization rely on superconducting platforms where proximity-induced pairing and spin–orbit coupling create topological superconducting states, recent theoretical advances have revealed that strongly spin–orbit coupled Mott insulators can provide an alternative pathway \cite{rau2016spin}. 
Among these, layered honeycomb iridates such as $Li_2IrO_3$ are particularly appealing, as their dominant bond-dependent Kitaev interactions naturally support fractionalized excitations. 
This view addresses the realization of Majorana zero modes (MZMs)\cite{sarma2015majorana} through purely magnetic mechanisms, eliminating the need for superconductivity Quantum spin liquids (QSLs)\cite{savary2016quantum, clark2021quantum} represent one of the most unconventional phases of quantum matter, lacking long-range magnetic order even at zero temperature \cite{balents2010spin}. 
At the same time, research on charge transport in low-dimensional materials such as Ge-doped phosphorene nanoribbons \cite{azizi2018charge} and graphene-oxide heterojunctions \cite{ghavami2015varistor} has shown nonlinear and quantum-coherent behaviors.  These results highlight the significance of nanoscale transport events for understanding and realizing topological excitations like Majorana fermions. 
Their ground states are characterized by long-range quantum entanglement, a feature not describable by conventional order parameters \cite{kitaev2006topological}. The Kitaev honeycomb model stands out as an exactly solvable example, where spin-$\frac{1}{2}$ moments fractionalize into itinerant Majorana fermions and static Z$_2$ gauge fields. This division confirms their resistance to local perturbations and their potential for topological quantum computing.  Moreover, when a magnetic field is applied, the Kitaev model can enter a gapped non-Abelian phase, evidenced experimentally by a half-quantized thermal Hall plateau\cite{kasahara2018majorana}.

\begin{figure}[htp]
\centering
\includegraphics[scale=0.135]{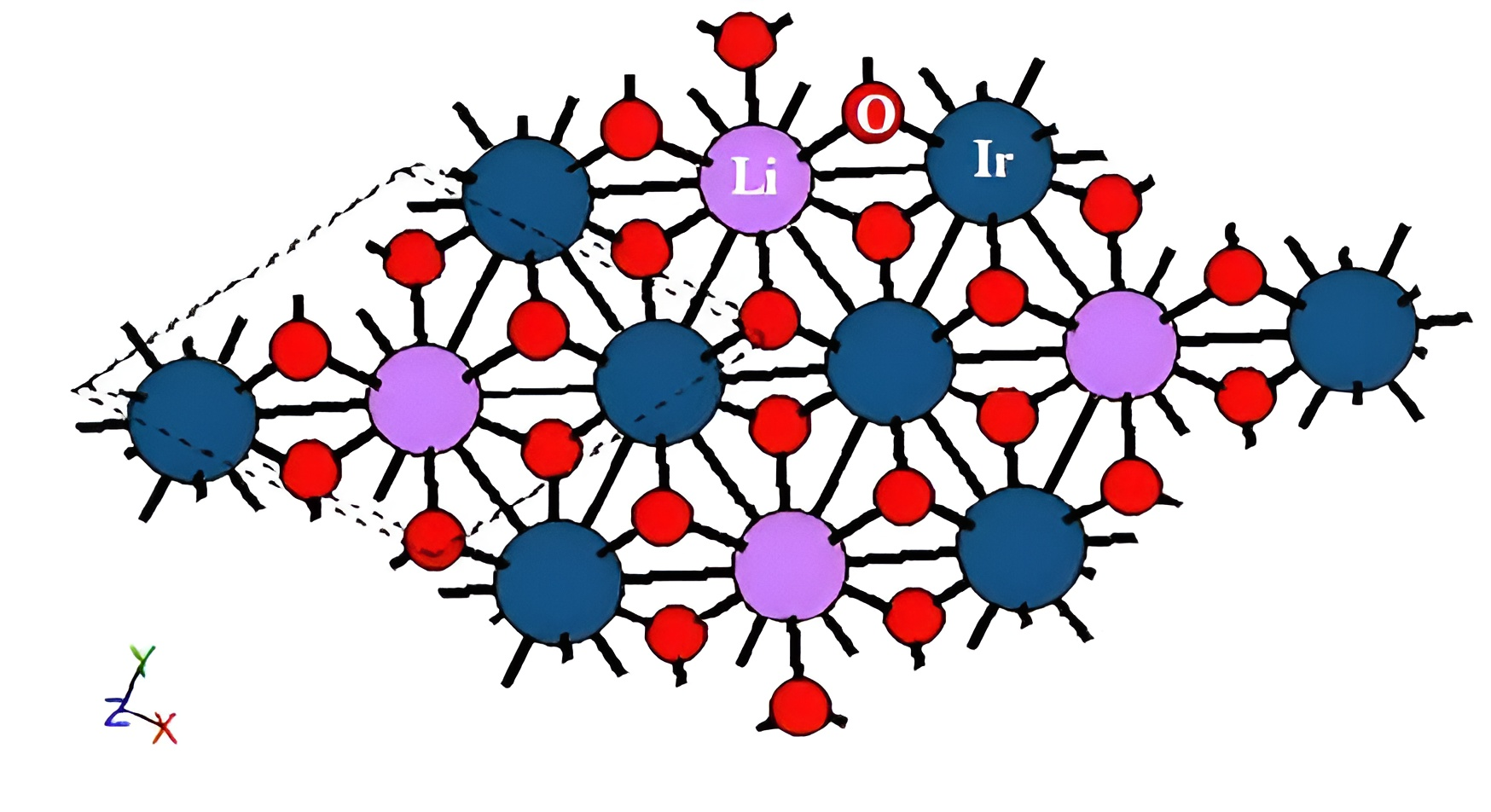}
\caption{Crystal structure of $Li_2IrO_3$. Navy blue spheres: Ir, red spheres: O, purple spheres: Li. Dashed lines indicate bonds behind the plane.}
\label{fig1}
\end{figure}

Strong spin–orbit coupled Mott insulators, including certain iridates and ruthenates, embody the key ingredients of the Kitaev model \cite{winter2017models,trebst2017kitaev}. In these systems, the interplay of electron correlations and relativistic spin–orbit coupling leads to highly anisotropic bond-dependent exchange, while geometric frustration suppresses conventional ordering. $Li_2IrO_3$ exemplifies this behavior: its edge-sharing $IrO_6$ octahedra generate bond-directional interactions, and its layered honeycomb structure fosters magnetic frustration Fig. \ref{fig1}. 
Experimental studies have reported features consistent with proximity to a Kitaev QSL, and theoretical work suggests that lattice defects, domain walls, or vortex-like spin textures in such a phase can bind MZMs without any superconducting proximity effect.
In this work, we investigate the emergence of MZMs in $Li_2IrO_3$ within the Kitaev–Heisenberg framework, focusing exclusively on magnetic interactions. By systematically varying the relative strengths of Kitaev and Heisenberg terms, we map out the regimes that favor fractionalization and stabilize topologically nontrivial excitations. To this end, we combine analytical spin–fractionalization techniques with numerical diagonalization of finite-size honeycomb clusters, allowing us to directly probe the Majorana sector of the theory. This approach provides a microscopic understanding of how a purely insulating, spin–orbit coupled magnet can emulate the essential features of topological superconductors, thereby expanding the landscape of candidate materials for fault-tolerant quantum computation.

\section{\label{sec:comp}THEORETICAL MODEL}
\begin{figure}[htp]
	\centering
	\begin{tikzpicture}[scale=2.5]
		\coordinate (A) at (0,0);
		\coordinate (B) at (1,0.5);
		\coordinate (C) at (0.5,1);
		\coordinate (D) at (-0.5,1.5);
		
		\draw[thick, red] (A) -- node[below,sloped] {$\gamma=x$} (B);
		\draw[thick, blue] (B) -- node[above,sloped] {$\gamma=y$} (C);
		\draw[thick, green!70!black] (C) -- node[below,sloped] {$\gamma=z$} (D);
		
		\foreach \p in {A,B,C,D}
		\draw[->, thick] (\p) -- ++(0.3,0.3) node[above] {$\mathbf{S}$};
		
		\node[red,right] at (2,1) {$K$-bond (x)};
		\node[blue,right] at (2,0.7) {$K$-bond (y)"};
		\node[green!70!black,right] at (2,0.4) {$K$-bond (z)"};
	\end{tikzpicture}
	\caption{(a) Kitaev-Heisenberg interactions in Li\textsubscript{2}IrO\textsubscript{3}'s hyperhoneycomb lattice. 
		(b) Bond-dependent spin components ($S^\gamma$) are shown for different nearest-neighbor links.}
	\label{fig:KH_Hamiltonian}
\end{figure}
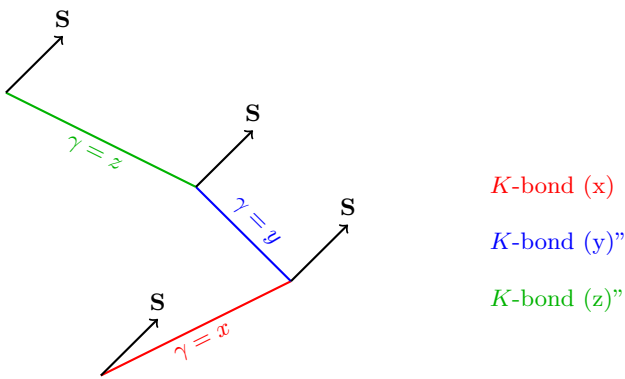

$Li_2IrO_3$ occupies a central place in the exploration of quantum magnetism, largely because it provides an experimental setting closely aligned with the Kitaev QSL framework. In this compound, the iridium ions form a honeycomb network, a geometry that, together with strong spin-orbit coupling, promotes highly anisotropic magnetic exchange. Such interactions, as described by the Kitaev model, can generate a magnetically disordered yet strongly entangled ground state in which conventional long-range order is absent. The interaction of these bond-dependent exchanges with additional magnetic couplings in $Li_2IrO_3$ has attracted extensive research attention, as it offers a rare opportunity to investigate Kitaev-type physics in a real material. Beyond its importance for understanding QSLs, this system is also considered a potential platform for future quantum information applications. Within a minimal theoretical description, its magnetic properties are captured by the Kitaev-Heisenberg Hamiltonian, which for $Li_2IrO_3$  takes the form:
\begin{equation}\label{H}
	\mathcal{H}=\sum_{<i,j>,\gamma}[\mathcal{J}_1 S_i.S_j+\mathcal{K}S_i^\gamma S_j^\gamma]+\sum_{<<i,j>>}\mathcal{J}_2 S_i.S_j
	\end{equation}

where $\mathcal{J}_1$, $\mathcal{K}$, and $\mathcal{J}_2$ are  Heisenberg exchange interaction\cite{} ($\thicksim 5$ MeV in $Li_2IrO_3$), Kitaev interaction\cite{takagi2019concept}($\mathcal{K}/\mathcal{J}_1 \approx -0.8 $  for iridates), and second-neighbor Heisenberg term receptivity, and $\gamma\in \{x,y,z\}$  is bond-dependent spin components, also in the equation \ref{H},   $S_i.S_j=(S^2-S_i^2-S_j^2)/2$. In order to be in k-space, we consider the Bloch Hamiltonian and apply the Majorana property. For simplicity, we consider the lattice constant a=1. Fig. \ref{fig:KH_Hamiltonian} shows the Heisenberg and Kitaev interactions in $Li_2IrO_3$’s hyperhoneycomb lattice.

The finite temperature dynamic polarization as a function of frequency and wave vector is defined as follows\cite{rostampour2024dynamic}
\begin{equation}
\Pi_q(\omega)=g\int\frac{d^3k}{(2\pi)^3}\frac{n_f(\epsilon_{k+q})-n_f(\epsilon_k)}{\omega+i\eta +\epsilon_k-\epsilon_{k+q}}.
\end{equation}
The Fermi-Dirac distribution function is mathematically represented as $n_f (\epsilon_k )=1/(e^{\beta\epsilon_k}+1)$, which represents the probability of occupying energy states by fermions in thermal equilibrium. Here, $\epsilon_k$ denotes the eigenvalues of the Hamiltonian, which indicate the energy levels associated with a quantum system. $\eta$ is an infinitesimal number. The Green's function is equal to
\begin{equation}
\hat{G}_k(i\omega)=[i\omega+\mu-H]^{-1}
\end{equation}
where the chemical potential is denoted by $\mu$. To study the electronic properties of $Li_2IrO_3$, the electron-electron interactions are systematically modeled through self-energy using the Green's function formalism. It is possible to analyze the behavior of the material at Matsubara frequencies. The self-energy is equal to
\begin{equation}
\hat{\Sigma}_k(i\omega_n)=\frac{1}{\beta}\sum_q\sum_{m=-\infty}^{\infty}\mathcal{W}_q(i\omega_m)\hat{G}_{k-q}(i\omega_n-i\omega_m)
\end{equation}
where the Matsubara frequencies are $\omega_n=\frac{2n\pi}{\beta}$, $\omega_m=\frac{(2m+1)\pi}{\beta}$\cite{chubukov2012first}. The q's are taken as $\vec{q}_1=\vec{k}-\vec{q}$, $\vec{q}_2=\mathcal{R}(\pi+\theta_{\vec{k}})\vec{q}_1$, and $\vec{q}_3=\mathcal{R}(\pi+\theta_{\vec{k}})\vec{q}_2$ such that $\mathcal{R}(\pi+\theta_{\vec{k}})$ is the rotation matrix with angle $(\pi+\theta_{\vec{k}})$. This research emphasizes the importance of considering screening effects to understand the complex behaviors exhibited by strongly correlated electron systems such as $Li_2IrO_3$. Screening effects play an important role in understanding the complex behaviors exhibited by strongly correlated electron systems. Effective screening potential is equivalent to
\begin{equation}
\mathcal{W}_q(i\omega_n)=\frac{\mathcal{V}_q}{1+\mathcal{V}_q\Pi_q(\omega)}.
\end{equation}
In the context of condensed matter physics, the interaction between charged particles is often described by the bare Coulomb interaction, denoted as $\mathcal{V}_q=\frac{2\pi e^2}{\kappa q}$, which is Fourier transformed into momentum space for analytical convenience. The parameter $\kappa$ represents the dielectric constant of the medium, which plays an important role in determining how these interactions are modified by the presence of other charges or external fields. The pole of the Green's function is the existence of collective excitations known as plasmons, which are crucial for the advancement of technologies in fields such as quantum computing and nanotechnology. The plasmon dispersion relation is critically determined by the equation $1+\mathcal{V}_q\Pi_q=0$.

In the study of $Li_2IrO_3$, the spectral function is used to analyze the excitations present, which provides information about the density of states at different energy levels. The spectral function of $Li_2IrO_3$ is the imaginary part of the Green's function, which is used to describe the propagation of particles and excitations in quantum systems. The spectral function of $Li_2IrO_3$ is as follows:

\begin{equation}
\begin{split}
\mathcal{A}_k(\omega)  = -2\Im[G_k(\omega)] 
= 2\pi\delta(\omega-\epsilon_k-\Re[\Sigma_k(\omega)])\\ = 2\mathcal{Z}_k\delta(\omega-E_k).
\end{split}
\end{equation}
In theoretical physics, the equation $\omega-\epsilon_k-\Re[\Sigma_k(\omega)]=0$ captures the fundamental principles governing the behavior of various systems.
\begin{equation}
\mathcal{Z}_k=\frac{1}{1-\frac{\partial}{\partial\omega}\Re[\Sigma_k(\omega=E_k)]}.
\end{equation}
The effective mass is obtained from a self-consistent solution to the Dyson equation, which plays an important role in understanding the electronic properties of materials. The Dyson equation, which describes the relationship between the Green's function of a system and its self-energy, allows for the inclusion of many-particle effects in the calculation of electronic states. The effective mass {has a significant role in determining the transport properties and band structures of semiconductors and other materials, and contributes significantly to their conductivity and overall performance in electronic applications. Effective mass calculations are a valuable tool in condensed matter physics\cite{rostampour2024dynamic}
\begin{equation}
\frac{m^*}{m}=\frac{\mathcal{Z}_k^{-1}}{1+\frac{m}{k_f}\frac{\partial}{\partial k}\Re[\Sigma_k(\omega=E_k)]}.
\end{equation}

\section{Results and discussion}
We focus on a parameter regime in which the system possesses a gapless spin liquid phase with edge-localized zero modes. The numerically computed spectral functions show sharp zero-bias peaks at the edges consistent with the presence of MZMs. We investigate thermal stability and robustness against disorder. The $Li_2IrO_3$ Kitaev model is a model that is used to describe the magnetic properties of a class of compounds known as spin-orbit coupled Mott insulators. This model is particularly relevant for the description of iridate behavior, with the heavy spin-orbit coupling leading to a highly anisotropic nearest-neighbor exchange interaction between neighboring spins.
For the $Li_2IrO_3$ system, the Kitaev model predicts novel states of matter that arise from magnetic interactions, such as QSLs, with exotic properties that defy conventional magnetic ordering. Because $Li_2IrO_3$ has a triangular lattice geometry, these unconventional magnetic interactions can arise, and thus, the latter is a central material in investigating new phases of quantum magnetism. Moreover, the understanding gained from the $Li_2IrO_3$ Kitaev model will be useful in further enhancement of understanding correlated electron systems and is applicable for future purposes in spintronics and quantum computing. In this article, we discuss the ferromagnetic Kitaev ($\mathcal{K} < 0$) and antiferromagnetic ($\mathcal{J} > 0$) Heisenberg exchange\cite{scherer2014unconventional}. By changing the polarization between the input and output photons, information about the angular momentum transfer and thus about the nature of the created excitations can be accessed \cite{toschi2020study}. Polarization dependence of magnetic resonance X-ray diffraction intensity allows for direct measurement of magnetic moment orientation \cite{biffin2014noncoplanar}. We study the spin-1/2 Kitaev-Heisenberg model on the bilayer honeycomb lattice with honeycomb planes coupled together by Heisenberg interactions. We adopt $\hbar=1$ and $\mathcal{J}_2=1$ in our numerical computation. \\
Kitaev-Heisenberg model is an important theoretical model used to describe the magnetic properties of materials, particularly in compounds like $Li_2IrO_3$. The model entails the utilization of the real and imaginary parts of the peaks of polarization, which are crucial in assessing the magnetic properties and quantum states of materials. The Majorana Hamiltonian in this model describes the interaction between particles and their excitations within a quantum system and provides details on their emergent phenomena and topological characteristics. Through the analysis of the peaks in polarization, researchers can have better insight into how real and imaginary components contribute to the unique magnetic orders and cooperative effects in $Li_2IrO_3$, ultimately advancing knowledge on quantum materials and their applications.
The study of the Kitaev-Heisenberg model, as considered particularly in the context of the compound $Li_2IrO_3$, reveals significant findings on the nature of magnetic interactions and phase transition in quantum spin systems.\\
Using this model, the maxima of polarization may be characterized by real and imaginary parts, which are indicative of valuable information on physical processes behind the phenomenon. The real parts characterize the response of the system to externally introduced perturbations, which are described by observable values such as magnetic order intensity. Imaginary parts characterize processes of energy dissipation and instability of some magnetic states. Understanding of the relationship between the imaginary and real components further enhances our knowledge of the emergent behavior in frustrated magnetism, thereby enriching our overall understanding of quantum materials and their potential for application in future technologies. The paramagnetic signal, being a consequence of the polarization coefficient of the neutron scattering cross-section, is sensitive to anisotropy that is dependent on the bonds \cite{kim2025kitaev}.
By a clever choice of phonon mode, amplitude, and polarization, particular spin couplings are selectively amplified compared to others. The available experimental couplings are sufficiently large as to already produce significant variations in the magnetic coupling, and phonon polarization may serve as an additional tunable parameter to design chiral interactions. Nearest-neighbor models illustrate the effects of phonon mode selection and phonon polarization on different exchange mechanisms \cite{kornjavca2024tuning}. New terahertz spectroscopy research on $Na_2Co_2TeO_6$ as a function of magnetic field applied with changing polarizations of terahertz exhibits spin dynamics with various characteristics over magnetic fields 0–70 kOe, 70–100 kOe, and $>$ 100 kOe. While, for lower than 70 kOe and higher than 100 kOe well-defined magnetic excitations dominate the dynamics, in the intermediate regime there is sharp absorption profile and broad continuum in the longitudinal as well as transverse polarization channels in both applied field directions $H//a$ and $H// a^*$. Polarization-selective continuum in the intermediate phase is an indication of spin fluctuations of an underlying proximate QSL\cite{bera2023}. 
\begin{figure}[htp]
\centering
\includegraphics[scale=0.122]{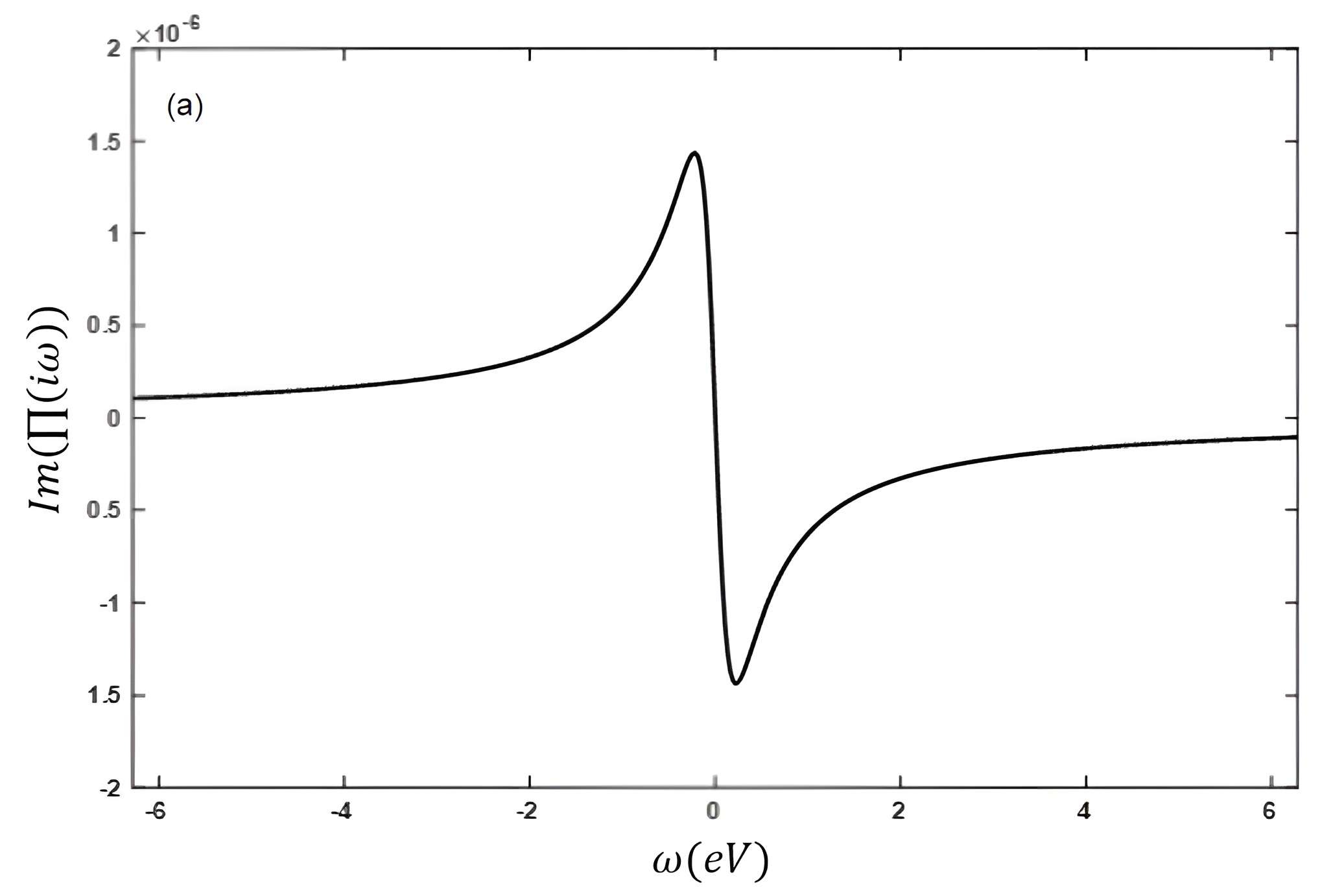}
\includegraphics[scale=0.122]{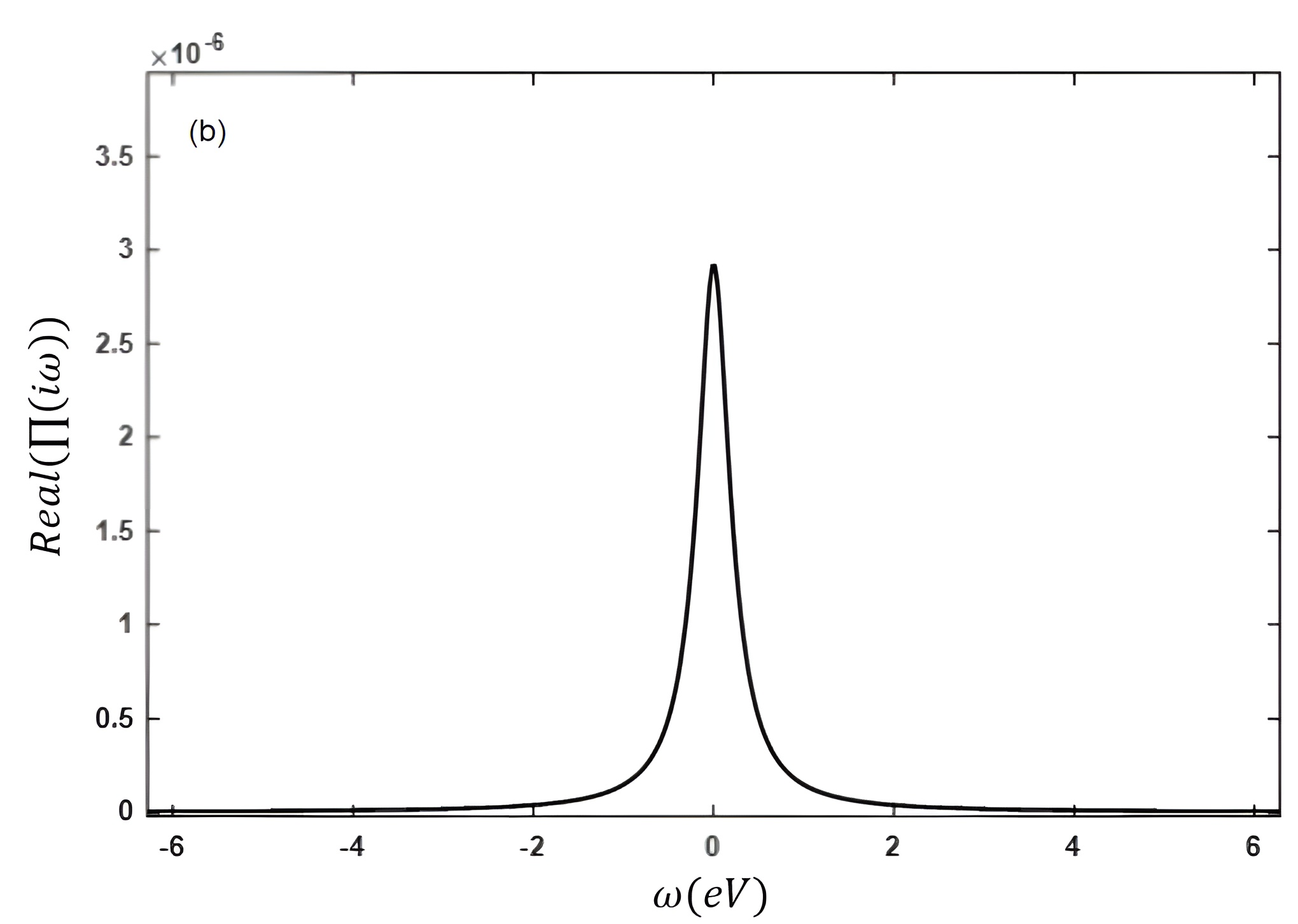}
\caption{ (Color online) Panel(a), imaginary of the polarization of the $Li_2IrO_3$ as a function of $\omega$ at T=10 K and for $q=0.2 nm^{-1}$ . Panel (b) the corresponding real part at the same conditions. }
\label{fig3}
\end{figure}
\begin{figure}[htp]
\centering
\includegraphics[scale=0.125]{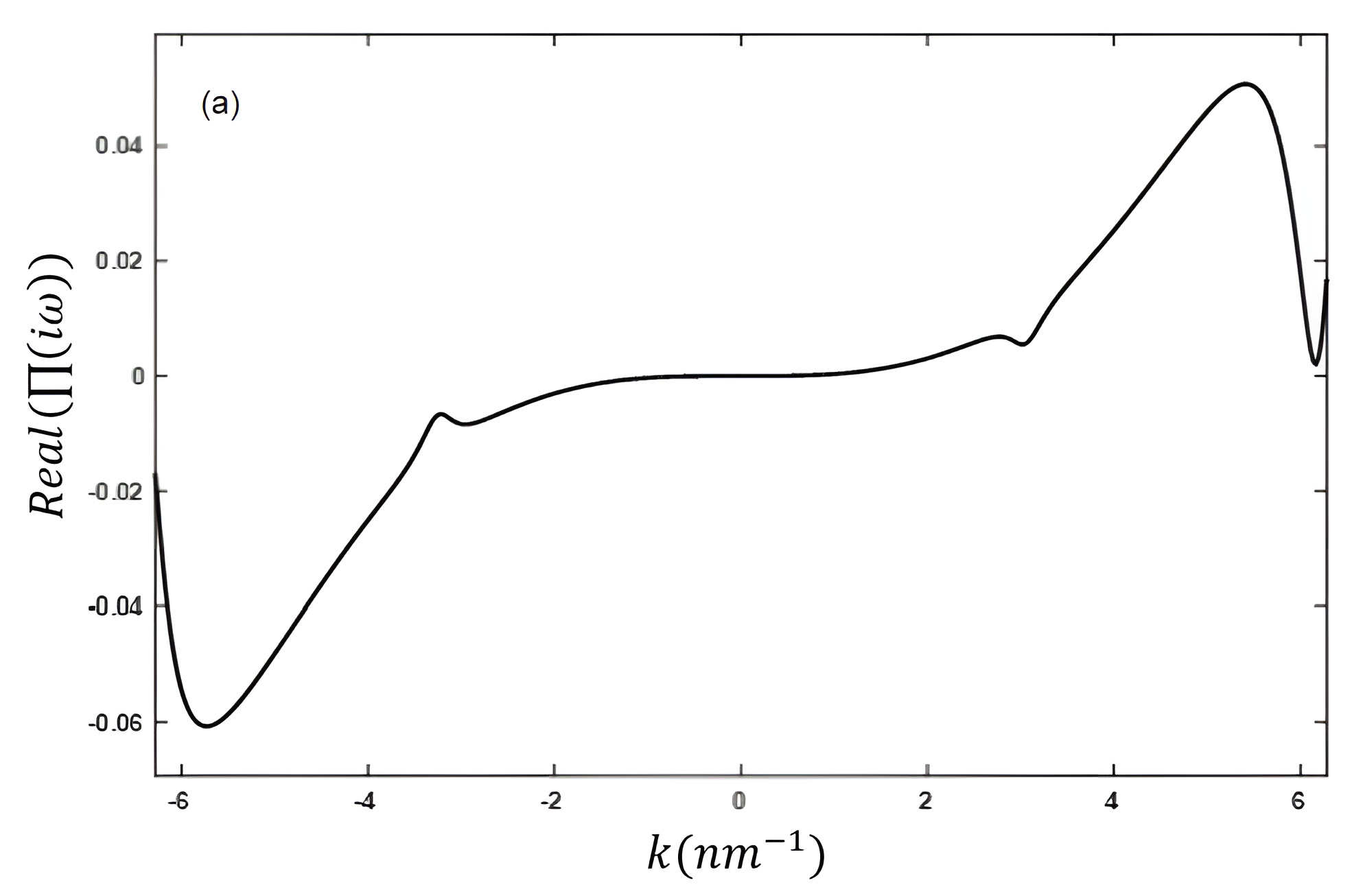}
\includegraphics[scale=0.125]{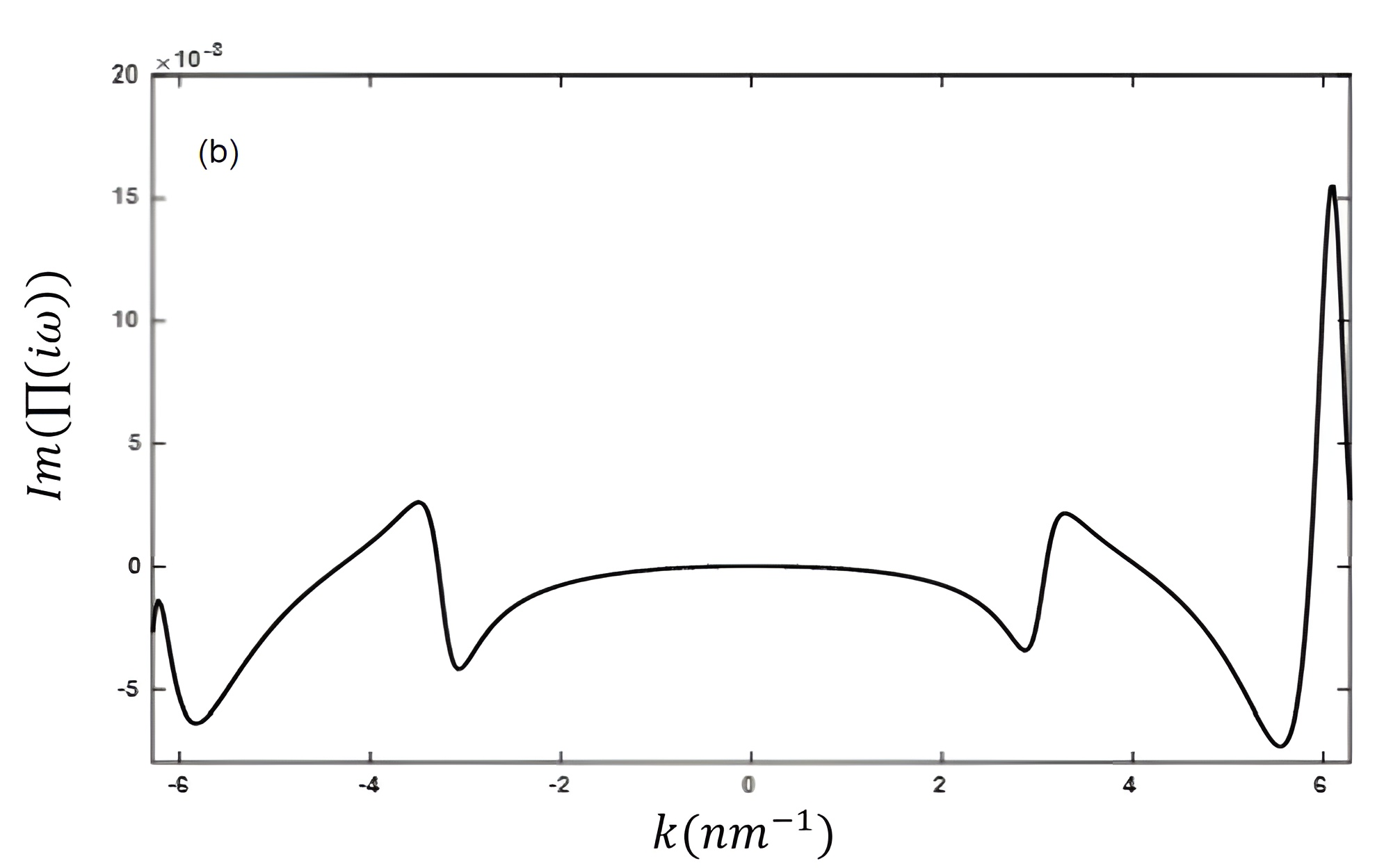}
\caption{(Color online) Panel(a), real part of the polarization of the $Li_2IrO_3$ as a function of  $k$ at T=10 K and for $q=0.2nm^{-1}$ . Panel (b) the corresponding imaginary part at the same conditions. }
\label{fig4}
\end{figure}
Fig. \ref{fig3} shows the imaginary and real parts of the polarization at 10 K vs. $\omega$. There is a sharp peak near the origin.
Fig. \ref{fig4} is a plot of the real and imaginary parts of the polarization as a function of the amplitude of the momentum $\mathcal{K}$. The fluctuations are larger than in Fig. \ref{fig3}. Theoretical and experimental studies have focused on a family of 4d and 5d tricoordinated compounds that are ostensibly close to the famed Kitaev model, one of very few exactly solvable models hosting gapless and gapped QSL ground states \cite{chen2008exact}. The magnetization data strongly confirm the dominance of the Kitaev-type ferromagnetic correlation and the vicinity of $\beta-Li_2IrO_3$ to the Kitaev spin liquid regime. 
Recent studies have established that $\beta-Li_2IrO_3$ is a highly promising candidate for the long-sought Kitaev spin liquid. Its anomalous magnetic properties result from its geometrical arrangement and strong spin-orbit coupling, which conspire to induce frustrated magnetic interactions. These results render 
$\beta-Li_2IrO_3$ a valuable focus of ongoing research on QSLs, with implications for advancing our understanding of quantum magnetism.
With such fascinating properties, $\beta-Li_2IrO_3$ can pave the way for future studies and applications in quantum computing and other fields, a milestone in condensed matter physics. It is possible that the presence of other interactions, finite but small, marginally stabilizes a non-collinear order below $T_c = 38 K$ \cite{takayama2015hyperhoneycomb}. The spectral function of the Kitaev-Heisenberg model is relevant to the investigation of the behaviors and properties of quantum spin systems. This model combines the Kitaev interactions, which are notorious for supporting Majorana fermions and topological properties, with the Heisenberg interactions that drive magnetic ordering. The spectral function tells us about the excitation spectrum of the system, including how the energy levels are occupied and how they evolve under various parameters like temperature and external magnetic fields.
Spectral function investigation is beneficial to researchers in examining the dynamics of quantum states, phase transitions, and emergent properties in these complex systems and is therefore a useful tool in theoretical and experimental condensed matter physics. Peaks in the spectral function in the Kitaev-Heisenberg model are crucial for comprehending the complex behaviors of quantum spin systems. It combines the Kitaev interaction, which creates strong frustration and leads to unconventional quantum states, with the Heisenberg interaction, which describes conventional magnetic coupling. The spectral function provides information on the energy distribution of excitations in the system, which reveals important features such as quasi-particle peaks and the presence of excitonic states. A study of these peaks in the spectral functions can also give further insights into the behavior of spin liquid phases and other emergent phenomena of quantum materials. The ongoing work on this subject continues to keep the interplay between all the interactions in the Kitaev-Heisenberg model at center stage, with the ability to lead to new breakthroughs in quantum magnetism and associated research fields. The spectral function comes into play to describe the density of states of a system across different energy levels. The spectral function is the Fourier transform of the regressive Green's function. The spectral function's peaks you see in the figure are energy levels at which particles such as electrons would be found to be.
Studying the spectral function assists scientists in probing quantum state dynamics, phase transitions, and emergent phenomena in such complex systems and hence constitutes an effective tool for theoretical and experimental condensed matter physics. Peaks in the spectral function of the Kitaev-Heisenberg model are also central to explaining the complex behavior of quantum spin systems. This model complements the Kitaev interaction, which builds up strong frustration and leads to exotic quantum states, and the Heisenberg interaction, which is behind traditional magnetic coupling. The spectral function provides information on excitation energy distribution in the system, introducing prominent features such as quasi-particle peaks and the presence of excitonic states. The research of such peaks of spectral functions can assist in further advancing the knowledge of the nature of spin liquid phases and other emergent quantities in quantum materials. As research continues, the interaction competition in the Kitaev-Heisenberg model is still a central question, which can have the potential to reveal fresh insight into quantum magnetism and related fields. The spectral function describes a system's density of states at different energy levels. The spectral function is the Fourier transform of the regressive Green's function. The peaks in the spectral function that you notice from the figure are energy levels where particles such as electrons are likely to be.
 \begin{figure}[htp]
\centering
\includegraphics[scale=0.125]{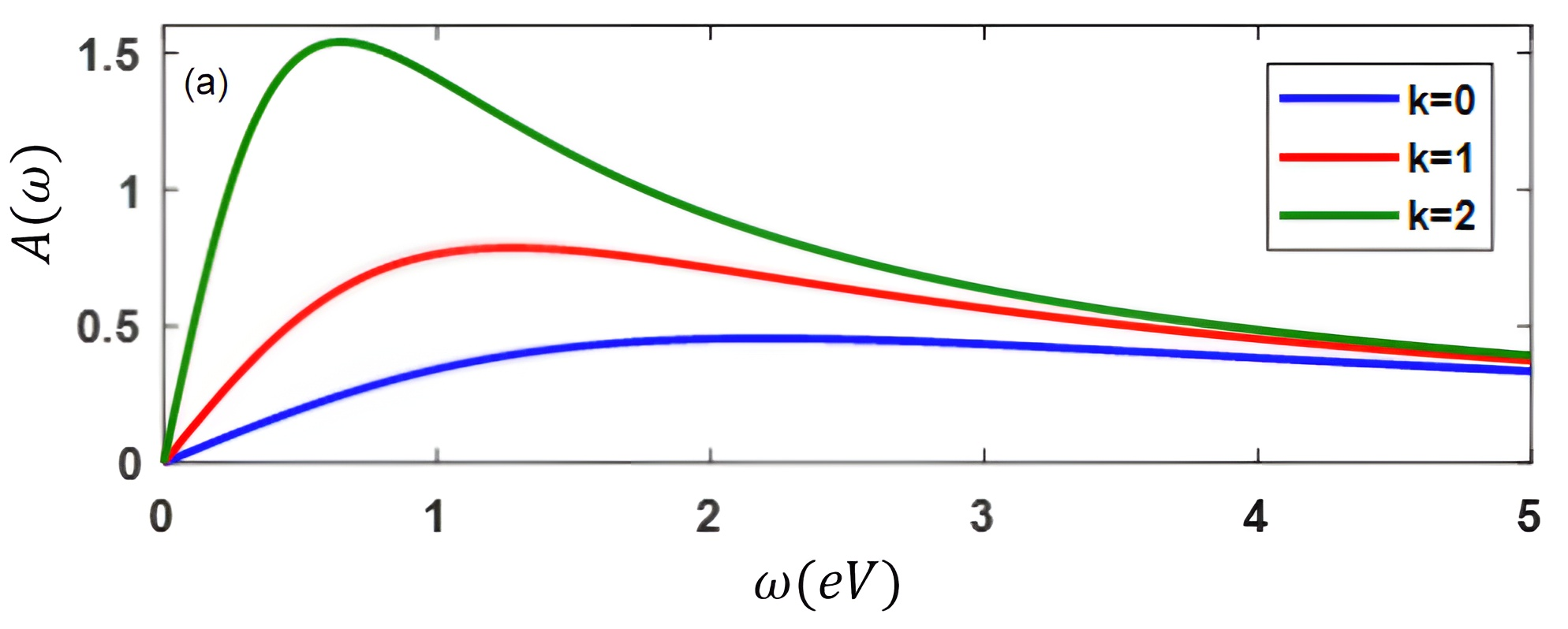}
\includegraphics[scale=0.15]{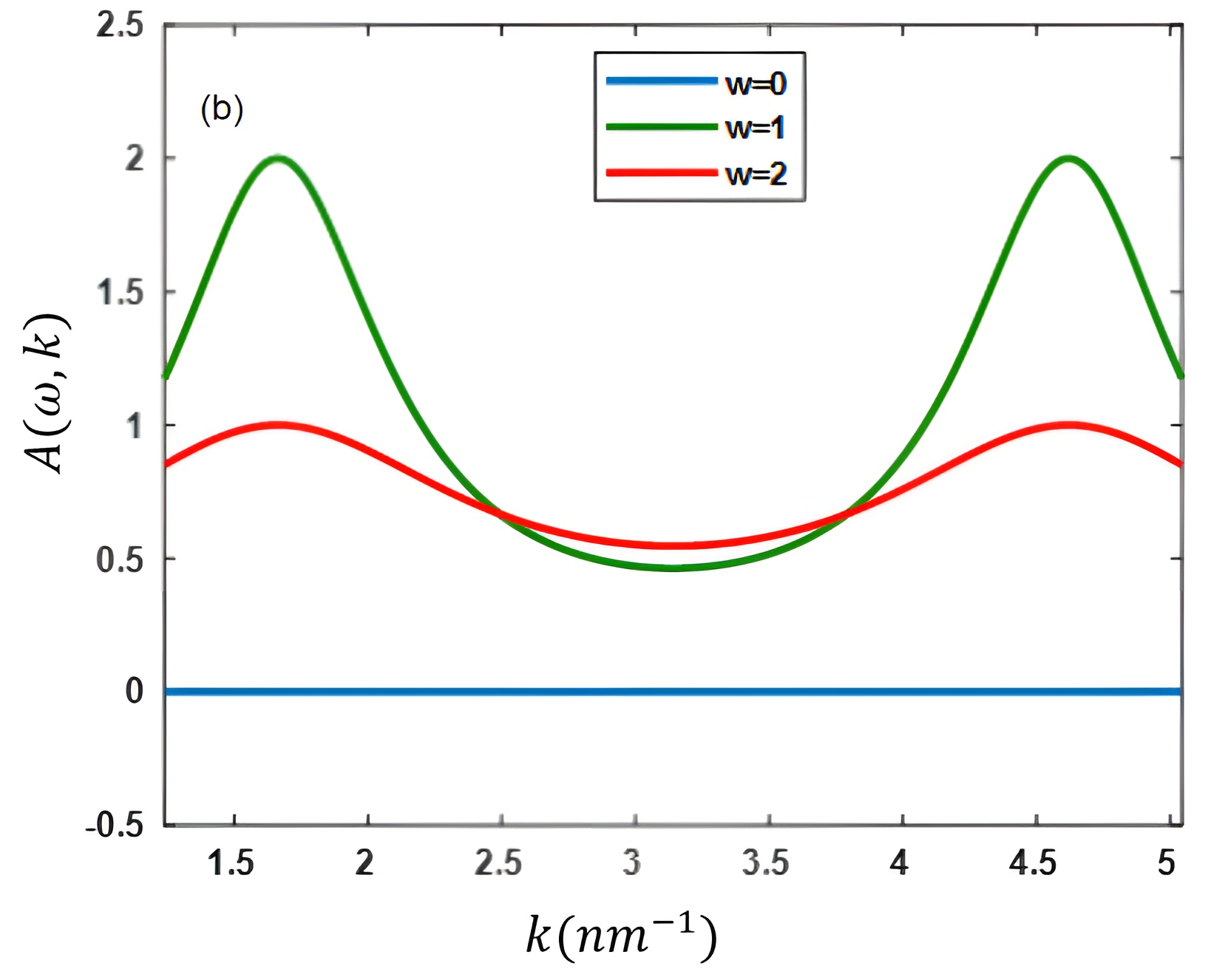}
\includegraphics[scale=0.157]{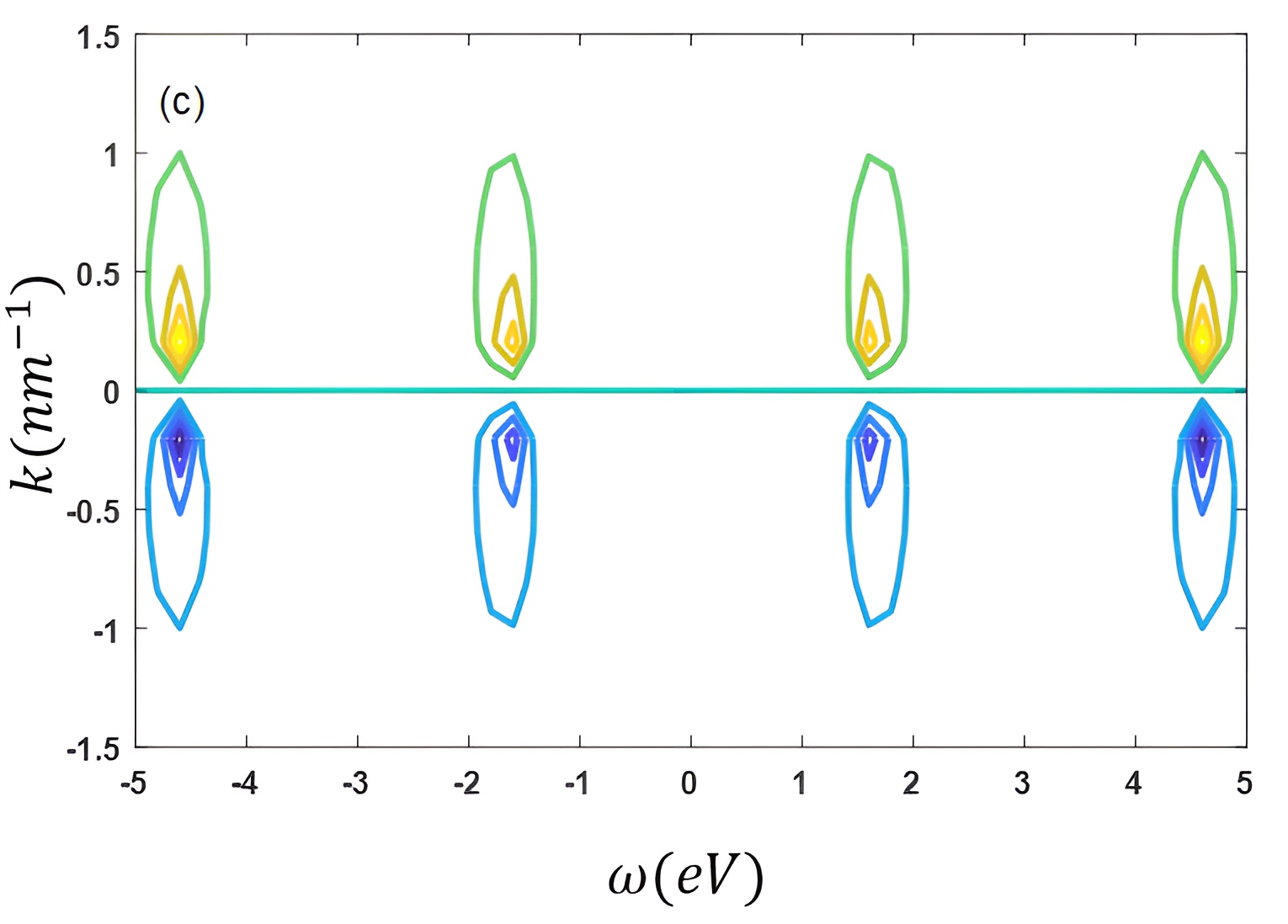}
\caption{(Color online) The spectral function of $Li_2IrO_3$.  The contour on the spectral function as a function of (a) $\omega$, (b) $k$, and (c) the contour along $k$ .}
\label{fig5}
\end{figure}
\begin{table*}[htp]
	\centering
	\begin{tabular}{|c|c|c|}
 \hline
		Application& Advantage of $Li_2IrO_3$& Description\\
  \hline
		 Quantum Spintronics & High SOC and no charge current & Efficient spin manipulation devices \\
   \hline
		 Magnetic Sensors & Topologically protected edge states & High sensitivity, low thermal noise \\
   \hline
		Topological Memory & Quantum spin liquid background & Long coherence times \\
  \hline
  Majorana Devices & Non-superconducting, scalable architecture & Reduced fabrication complexity \\
  \hline
	\end{tabular}
	\caption{Potential industrial applications of $Li_2IrO_3$ as a superconducting Majorana platform.}
	\label{tab:lithium_iridate}
\end{table*}
The spectral function as a function of the extent of the K and $\omega$ is shown in Fig. \ref{fig5}. From the figure, there is one peak versus $\omega$ and two peaks versus K, other than when $\omega=0$. The contour versus K is also plotted in Fig. \ref{fig5}.
Bin-Bin Wang et al.\cite{wang2018dynamics} showed that in the zigzag phase, coherent quasi-particle features are exhibited clearly in spectral functions in the first Brillouin zone for holes created and annihilated on a sublattice, but they are hidden in physical spectral functions of angularly resolved optical emission spectroscopy experiments such that these hidden spectral retrievals fall within extended Brillouin zones \cite{wang2018dynamics}. Dynamical hole spectral functions provide rich information about the structure of fractional QSLs \cite{kadow2024single}. The research revealed that the effective mass for fermionic quasi-particles like in $Ag_3LiIr_2O_6$ is, surprisingly, on the same order as the bare electron's effective mass \cite{heath2023signatures}. For the cuprate family, the effective mass increases with doping at approximately the same rate. This is consistent with the observed variation of effective mass in $(Sr_{1-x}La_x)_3Ir_2O_7$, which has a correspondence to the change in effective mass with doping in the iridates and this high-energy renormalization in the cuprates. This high effective mass in the metallic samples is on the order of, though a little higher than, values in a recent work in which the increase in mass was implied indirectly from infrared spectroscopy \cite{ahn2016infrared, meevasana2007hierarchy}.
\begin{figure}[htp]
\centering
\includegraphics[scale=0.145]{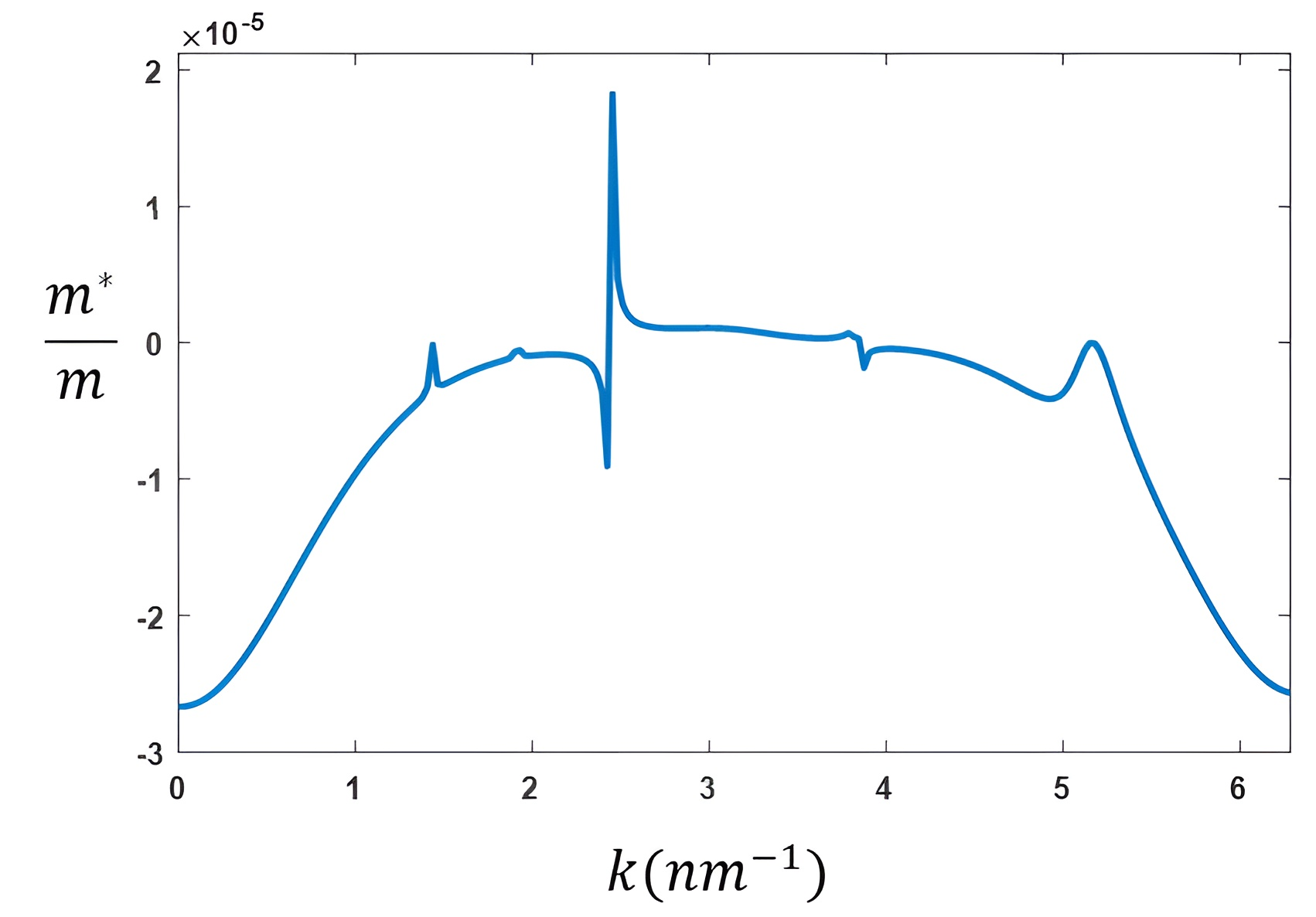}
\caption{(Color online) The effective mass of $Li_2IrO_3$ as function of $k$ with $q=5nm^{-1}1$  and T=10 K. }
\label{fig6}
\end{figure}
The effective mass is graphed in Fig. \ref{fig6} versus K, with its peak occurring at K=2.4 $nm^{-1}$. The charge carriers at the maximum of the effective mass have greater mobility, which leads to superior electrical conductivity and enhanced performance in semiconductor devices. Effective mass is the mass an electron or a hole effectively possesses when it reacts to forces, such as electric or magnetic fields, in a solid. The Kitaev-Heisenberg model's effective mass peak is a significant advancement in the quantum magnetism research, particularly for spin systems. The model adds the Kitaev interaction, which has the bond-dependent interactions and strong spin-orbit coupling, and the isotropic Heisenberg exchange interaction, which includes the spin-spin coupling. The maximum of effective mass in this description is a remarkable point at which the character of the spin system alters, reflecting features of state degeneracy and topological order. Physicists use this description to investigate a broad range of quantum material phenomena involving quantum phase transitions and emergent excitations and yielding insights that can lead to new quantum computing and spintronic device applications. \\
\section{Industrial application}
$Li_2IrO_3$ is a compound that exhibits fascinating magnetic and electronic properties, which makes it a hot subject in condensed matter physics. The Kitaev-Heisenberg model, when taken in the context of the material $Li_2IrO_3$, gives significant industrial implications in the field of quantum computing as well as spintronics. The model with both Kitaev interactions and Heisenberg exchange leads to fascinating magnetic properties and exotic phases, e.g., Majorana-bound states. $Li_2IrO_3$, a prime candidate in realizing such frameworks due to its remarkable electronic structure and high spin-orbit coupling, shows evidence of breaks in realizing quantum devices. The understanding and manipulation of Majorana modes from this Hamiltonian would pave the way towards fault-tolerant quantum computation because Majorana fermions are anticipated to be immune to external perturbations. Through investigating these industrial uses, scientists are not just pushing the limits of basic physics but also paving the way for future technologies that leverage quantum mechanics for everyday applications. Table \ref{tab:lithium_iridate} presents its industrial uses.\\
\section{Conlclusion}
Our study puts $Li_2IrO_3$ on a realistic list of materials for the realization of superconducting Majorana fermions. The layer structure, spin-orbit coupling, and Kitaev interactions in combination offer a novel platform for future quantum technologies. Our proposal outlines the theoretical feasibility and industrial relevance of such systems.
The real and imaginary parts of the polarization peaks play an important role in evaluating the magnetic properties and quantum states of the materials. Numerical results showed that the sharp peak in the spectral functions indicates zero bias at the edges. The Kitaev-Heisenberg model is used to describe the magnetic properties. The geometric arrangement and strong spin-orbit coupling affect the magnetic properties of $Li_2IrO_3$. The spectral function shows quasi-particle peaks and the presence of excitonic states. The results indicate that the charge carriers at the maximum effective mass have higher mobility.\\

\section{Declaration of Interest}
The authors declare that they have no known competing financial interests or personal relationships that could have appeared to influence the work reported in this paper.

\bibliography{ref}

@article{beenakker2013search,
  title={Search for Majorana fermions in superconductors},
  author={Beenakker, Carlo WJ},
  journal={Annu. Rev. Condens. Matter Phys.},
  volume={4},
  number={1},
  pages={113--136},
  year={2013},
  publisher={Annual Reviews}
}

@article{leijnse2012introduction,
  title={Introduction to topological superconductivity and Majorana fermions},
  author={Leijnse, Martin and Flensberg, Karsten},
  journal={Semiconductor Science and Technology},
  volume={27},
  number={12},
  pages={124003},
  year={2012},
  publisher={IOP Publishing}
}

@article{stanescu2013majorana,
  title={Majorana fermions in semiconductor nanowires: fundamentals, modeling, and experiment},
  author={Stanescu, Tudor D and Tewari, Sumanta},
  journal={Journal of Physics: Condensed Matter},
  volume={25},
  number={23},
  pages={233201},
  year={2013},
  publisher={IOP Publishing}
}

@article{kozii2016three,
  title={Three-dimensional Majorana fermions in chiral superconductors},
  author={Kozii, Vladyslav and Venderbos, J{\"o}rn WF and Fu, Liang},
  journal={Science advances},
  volume={2},
  number={12},
  pages={e1601835},
  year={2016},
  publisher={American Association for the Advancement of Science}
}

@article{rau2016spin,
  title={Spin-orbit physics giving rise to novel phases in correlated systems: Iridates and related materials},
  author={Rau, Jeffrey G and Lee, Eric Kin-Ho and Kee, Hae-Young},
  journal={Annual Review of Condensed Matter Physics},
  volume={7},
  number={1},
  pages={195--221},
  year={2016},
  publisher={Annual Reviews}
}

@article{savary2016quantum,
  title={Quantum spin liquids: a review},
  author={Savary, Lucile and Balents, Leon},
  journal={Reports on Progress in Physics},
  volume={80},
  number={1},
  pages={016502},
  year={2016},
  publisher={IOP Publishing}
}

@article{clark2021quantum,
  title={Quantum spin liquids from a materials perspective},
  author={Clark, Lucy and Abdeldaim, Aly H},
  journal={Annual Review of Materials Research},
  volume={51},
  number={1},
  pages={495--519},
  year={2021},
  publisher={Annual Reviews}
}

@article{sarma2015majorana,
  title={Majorana zero modes and topological quantum computation},
  author={Sarma, Sankar Das and Freedman, Michael and Nayak, Chetan},
  journal={npj Quantum Information},
  volume={1},
  number={1},
  pages={1--13},
  year={2015},
  publisher={Nature Publishing Group}
}

@article{balents2010spin,
  title={Spin liquids in frustrated magnets},
  author={Balents, Leon},
  journal={nature},
  volume={464},
  number={7286},
  pages={199--208},
  year={2010},
  publisher={Nature Publishing Group UK London}
}

@article{kasahara2018majorana,
  title={Majorana quantization and half-integer thermal quantum Hall effect in a Kitaev spin liquid},
  author={Kasahara, Yuichi and Ohnishi, Tsuneya and Mizukami, Yuta and Tanaka, Osamu and Ma, Sixiao and Sugii, Kaori and Kurita, Nobuyuki and Tanaka, Hidekazu and Nasu, Joji and Motome, Yukitoshi and others},
  journal={Nature},
  volume={559},
  number={7713},
  pages={227--231},
  year={2018},
  publisher={Nature Publishing Group UK London}
}

@article{trebst2017kitaev,
  title={Kitaev materials},
  author={Trebst, Simon},
  journal={arXiv preprint arXiv:1701.07056},
  year={2017}
}

@article{rostampour2024dynamic,
  title={Dynamic polarization, quantum spectral function and effective mass for black phosphorous: Random phase approximation approach at finite temperature},
  author={Rostampour, Elnaz and Ghavami, Badie and Herrera, Saul A and Naumis, Gerardo G},
  journal={Solid State Communications},
  volume={384},
  pages={115497},
  year={2024},
  publisher={Elsevier}
}

@article{winter2017models,
  title        = {Models and materials for generalized Kitaev magnetism},
  author       = {Winter, Stephen M. and Tsirlin, Alexander A. and Daghofer, Maria and van den Brink, Jeroen and Singh, Yogesh and Gegenwart, Philipp and Valenti, Roser},
  journal      = {J. Phys.: Condens. Matter},
  volume       = {29},
  pages        = {493002},
  year         = {2017},
  doi          = {10.1088/1361-648X/aa8cf5}
}

@article{bera2023,
  author = {Bera, A. K. and Yusuf, S. M. and Orlandi, F. and Manuel, P. and Bhaskaran, L. and Zvyagin, S. A.},
  title = {Magnetic excitations in the Kitaev candidate {Na$_2$Co$_2$TeO$_6$} probed by inelastic neutron scattering},
  journal = {Phys. Rev. B},
  volume = {108},
  pages = {214419},
  year = {2023}
}

@article{scherer2014unconventional,
  title={Unconventional pairing and electronic dimerization instabilities in the doped Kitaev-Heisenberg model},
  author={Scherer, Daniel D and Scherer, Michael M and Khaliullin, Giniyat and Honerkamp, Carsten and Rosenow, Bernd},
  journal={Physical Review B},
  volume={90},
  number={4},
  pages={045135},
  year={2014},
  publisher={APS}
}

@article{toschi2020study,
  title={Study of low-energy excitations in the pyrochlore iridate Tb (2+ x) Ir (2-x) O (7-y)(x= 0.4) probed by resonant inelastic X-ray scattering},
  author={Toschi, Anna},
  year={2020}
}

@article{biffin2014noncoplanar,
  title={Noncoplanar and counterrotating incommensurate magnetic order stabilized by Kitaev interactions in $\gamma$-Li 2 IrO 3},
  author={Biffin, A and Johnson, RD and Kimchi, I and Morris, R and Bombardi, A and Analytis, JG and Vishwanath, A and Coldea, R},
  journal={Physical review letters},
  volume={113},
  number={19},
  pages={197201},
  year={2014},
  publisher={APS}
}

@article{kim2025kitaev,
  title={Kitaev interaction and proximate higher-order skyrmion crystal in the triangular lattice van der Waals antiferromagnet NiI2},
  author={Kim, Chaebin and Vilella, Olivia and Lee, Youjin and Park, Pyeongjae and An, Yeochan and Cho, Woonghee and Stone, Matthew B and Kolesnikov, Alexander I and Hao, Yiquing and Asai, Shinichiro and others},
  journal={arXiv preprint arXiv:2502.14167},
  year={2025}
}

@article{kornjavca2024tuning,
  title={Tuning magnetic interactions with nonequilibrium optical phonon populations},
  author={Kornja{\v{c}}a, Milan and Flint, Rebecca},
  journal={arXiv preprint arXiv:2410.21373},
  year={2024}
}

@article{chen2008exact,
  title={Exact results of the Kitaev model on a hexagonal lattice: spin states, string and brane correlators, and anyonic excitations},
  author={Chen, Han-Dong and Nussinov, Zohar},
  journal={Journal of Physics A: Mathematical and Theoretical},
  volume={41},
  number={7},
  pages={075001},
  year={2008},
  publisher={IOP Publishing}
}

@article{takayama2015hyperhoneycomb,
  title={Hyperhoneycomb iridate $\beta$-Li 2 IrO 3 as a platform for Kitaev magnetism},
  author={Takayama, Tomohiro and Kato, Akihiko and Dinnebier, Robert and Nuss, J{\"u}rgen and Kono, H and Veiga, LSI and Fabbris, G and Haskel, D and Takagi, H},
  journal={Physical review letters},
  volume={114},
  number={7},
  pages={077202},
  year={2015},
  publisher={APS}
}

@article{wang2018dynamics,
  title={Dynamics of a single hole in the Heisenberg--Kitaev model: a self-consistent Born approximation study},
  author={Wang, Bin-Bin and Wang, Wei and Yu, Shun-Li and Li, Jian-Xin},
  journal={Journal of Physics: Condensed Matter},
  volume={30},
  number={38},
  pages={385602},
  year={2018},
  publisher={IOP Publishing}
}

@article{kadow2024single,
  title={Single-hole spectra of Kitaev spin liquids: from dynamical Nagaoka ferromagnetism to spin-hole fractionalization},
  author={Kadow, Wilhelm and Jin, Hui-Ke and Knolle, Johannes and Knap, Michael},
  journal={npj Quantum Materials},
  volume={9},
  number={1},
  pages={32},
  year={2024},
  publisher={Nature Publishing Group UK London}
}

@article{heath2023signatures,
  title={Signatures of a Majorana-Fermi surface in the Kitaev magnet Ag3LiIr2O6},
  author={Heath, Joshuah T and Bahrami, Faranak and Lee, Sangyun and Movshovich, Roman and Chen, Xiao and Tafti, Fazel and Bedell, Kevin},
  journal={Communications Physics},
  volume={6},
  number={1},
  pages={348},
  year={2023},
  publisher={Nature Publishing Group UK London}
}

@article{ahn2016infrared,
  title={Infrared Spectroscopic Evidences of Strong Electronic Correlations in (Sr1- x La x) 3Ir2O7},
  author={Ahn, Gihyeon and Song, SJ and Hogan, T and Wilson, SD and Moon, SJ},
  journal={Scientific reports},
  volume={6},
  number={1},
  pages={32632},
  year={2016},
  publisher={Nature Publishing Group UK London}
}

@article{meevasana2007hierarchy,
  title={Hierarchy of multiple many-body interaction scales in high-temperature superconductors},
  author={Meevasana, W and Zhou, XJ and Sahrakorpi, S and Lee, WS and Yang, WL and Tanaka, K and Mannella, N and Yoshida, T and Lu, DH and Chen, YL and others},
  journal={Physical Review B—Condensed Matter and Materials Physics},
  volume={75},
  number={17},
  pages={174506},
  year={2007},
  publisher={APS}
}

@article{kitaev2006topological,
  title={Topological entanglement entropy},
  author={Kitaev, Alexei and Preskill, John},
  journal={Physical review letters},
  volume={96},
  number={11},
  pages={110404},
  year={2006},
  publisher={APS}
}

@article{takagi2019concept,
  title={Concept and realization of Kitaev quantum spin liquids},
  author={Takagi, Hidenori and Takayama, Tomohiro and Jackeli, George and Khaliullin, Giniyat and Nagler, Stephen E},
  journal={Nature Reviews Physics},
  volume={1},
  number={4},
  pages={264--280},
  year={2019},
  publisher={Nature Publishing Group UK London}
}

@article{chubukov2012first,
  title={First-Matsubara-frequency rule in a Fermi liquid. I. Fermionic self-energy},
  author={Chubukov, Andrey V and Maslov, Dmitrii L},
  journal={Physical Review B—Condensed Matter and Materials Physics},
  volume={86},
  number={15},
  pages={155136},
  year={2012},
  publisher={APS}
}

@article{azizi2018charge,
  title={Charge transport in germanium doped phosphorene nanoribbons},
  author={Azizi, Maryam and Ghavami, Badie},
  journal={RSC advances},
  volume={8},
  number={35},
  pages={19479--19485},
  year={2018},
  publisher={Royal Society of Chemistry}
}

@article{ghavami2015varistor,
  title={Varistor characteristics of a nano-device containing graphene and oxidised graphene: verification by DFT+ NEGF},
  author={Ghavami, Badie and Rastkar-Ebrahimzadeh, Alireza},
  journal={Molecular Physics},
  volume={113},
  number={23},
  pages={3696--3702},
  year={2015},
  publisher={Taylor \& Francis}
}

\end{document}